\documentclass[sigconf,nonacm]{acmart}

\usepackage{amsmath}
\usepackage{cleveref}
\usepackage{listings}
\usepackage{enumitem}
\usepackage{glossaries}
\usepackage{multirow}
\usepackage{microtype}
\usepackage{url}

\graphicspath{{images/}}

\newacronym{cdn}{CDN}{content delivery network}
\newacronym{url}{URL}{uniform resource locator}
\newacronym{tpl}{TPL}{third-party library}
\newacronym{sca}{SCA}{software composition analysis}
\newacronym{ast}{AST}{abstract syntax tree}
\glsdisablehyper

\setcopyright{acmlicensed}
\copyrightyear{2026}
\acmYear{2026}
\acmDOI{XXXXXXX.XXXXXXX}
\acmConference[ICSE '26]{International Conference on Software Engineering%
  }{April 12--18, 2026}{Rio de Janeiro, Brazil}
\acmISBN{978-1-4503-XXXX-X/2026/04}

\begin{document}

\title{Insecure Ingredients? Exploring Dependency Update Patterns of Bundled JavaScript Packages on the Web}

\author{Ben Swierzy}
\email{swierzy@cs.uni-bonn.de}
\orcid{0009-0003-0485-4791}
\affiliation{%
  \institution{Fraunhofer FKIE}
  \institution{University of Bonn}
  \city{Bonn}
  \country{Germany}
}
\author{Marc Ohm}
\email{ohm@cs.uni-bonn.de}
\orcid{0000-0002-2913-5270}
\affiliation{%
  \institution{University of Bonn}
  \institution{Fraunhofer FKIE}
  \city{Bonn}
  \country{Germany}
}
\author{Michael Meier}
\email{mm@cs.uni-bonn.de}
\orcid{0009-0006-8199-5004}
\affiliation{%
  \institution{University of Bonn}
  \institution{Fraunhofer FKIE}
  \city{Bonn}
  \country{Germany}
}

\renewcommand{\shortauthors}{Swierzy et al.}

\begin{abstract}
    Reusable software components, typically distributed as packages, are a central paradigm of modern software development.
    The JavaScript ecosystem serves as a prime example, offering millions of packages with their use being promoted as idiomatic.
    However, download statistics on npm raise security concerns as they indicate a high popularity of vulnerable package versions while their real prevalence on production websites remains unknown. 
    Package version detection mechanisms fill this gap by extracting utilized packages and versions from observed artifacts on the web.
    Prior research focuses on mechanisms for either hand-selected popular packages in bundles or for single-file resources utilizing the global namespace.
    This does not allow for a thorough analysis of modern web applications' dependency update behavior at scale.
    In this work, we improve upon this by presenting Aletheia, a package-agnostic method which dissects JavaScript bundles to identify package versions through algorithms originating from the field of plagiarism detection.
    We show that Aletheia clearly outperforms the existing approaches in practical settings.
    Furthermore, we crawl the Tranco top 100,000 domains to reveal that 5\% -- 20\% of domains update their dependencies within 16 weeks.
    Surprisingly, from a longitudinal perspective, bundled packages are updated significantly faster than their CDN-included counterparts, with consequently up to 10 times fewer known vulnerable package versions included.
    Still, we observe indicators that few widespread vendors seem to be a major driving force behind timely updates, implying that quantitative measures are not painting a complete picture.
\end{abstract}




\maketitle

\section{Introduction}
\label{sec:introduction}

Reuse is a well-established concept in modern software development that has led to the ubiquity of software packages.
This is particularly evident in agile environments such as the web.
In this area, JavaScript is the dominant programming language for developing interactive and modern frontends.
Over the years, a large ecosystem emerged from the idiomatic view to utilize packages even for tasks with low complexity~\cite{Kula2017}.
This led to the software registry around the package manager npm becoming the largest of its kind, with its contents known for deep dependency chains.
Critical vulnerabilities like Log4Shell increased the global awareness and effort to improve the security of the software supply chain with a central security advice for end users~\cite{Redmiles2016} also applying for web application developers: Updates are important.

The GitHub security advisory database contains all CVE-num\-bered vulnerabilities that affect packages hosted on npm.
The amount of new entries maintains a constant high level each year, emphasizing the importance of an active dependency management.
Nonetheless, the deprecated \texttt{angular} package is a prime example hinting severe issues in the adoption:
It is downloaded more than 400,000 times every week of the past 12 months, despite eight known vulnerabilities with high and medium severity.
However, it has already been shown that production dependencies cannot be reliably inferred from lists of all dependencies~\cite{Latendresse2022}.
Similarly, download statistics cannot be regarded as a reliable measurement since they can also correspond to local development setups with vastly different security requirements compared to production servers.

To study the update behavior of JavaScript resources on the web and its implications for security, this paper proposes a methodology to detect package versions on websites.
The problem can be formulated as follows:
Given a website, we want to automatically determine the list of used packages with their versions.
There are three primary ways for web applications to incorporate packages.
First, they can be included using the hosting services of a public \gls{cdn}.
Since the corresponding \glspl{url} encode the package name and version, the problem is trivially solved.
Second, the domain may host the required packages itself, through an associated service or even inline them.
In this scenario, hash-based identification and global variables have been successfully applied in the past~\cite{Lauinger2017,Pagon2023,Stock2017,Demir2021}.
Third, multiple packages may be compiled into a single bundle, usually hosted on the domain itself.
This removes the need for determining the correct package inclusion order and reduces HTTP requests.
It is considered the more modern approach and is prevalent nowadays~\cite{Rack2023}.
In this scenario, there is no general and reliable way known yet to solve the problem.

In this paper, we aim to address this gap.
In particular, our contributions comprise (1) a method for reliable extraction of relevant source code from npm release artifacts, (2) an algorithm to detect the versions of arbitrary bundled packages, and (3) an analysis of the update behavior of JavaScript packages over the top 100,000 domains.
All scripts and artifacts are openly available at \url{https://doi.org/10.60507/FK2/F4VQRH} (scripts), \url{https://doi.org/10.60507/FK2/F4VQRH} (dolospy) and \url{https://doi.org/10.60507/FK2/AHYGMN} (data) to facilitate reproducibility and future research.
With these contributions, we answer the following research questions:

\begin{enumerate}[label=\bfseries RQ\arabic*\normalfont,left=0mm]
    \item How prevalent are JavaScript bundles and CDN inclusions?
    \item Which method enables reliable detection of package versions in JavaScript bundles?
    \item Do production websites deploy package updates timely, reducing the window of opportunity for known security vulnerabilities to be exploited? 
\end{enumerate}

The paper is organized as follows.
\Cref{sec:background} provides background knowledge on CDNs, bundle generation and bundle analysis approaches.
\Cref{sec:methodology} introduces our version detection algorithm.
\Cref{sec:evaluation} describes the evaluation methodology, while \Cref{sec:results} presents the results.
In \Cref{sec:discussion} we discuss results and limitations.
Related work and ethical aspects are considered in \Cref{sec:related-work} and \ref{sec:ethical-considerations}.
We conclude in \Cref{sec:conclusion}.

\section{Background}
\label{sec:background}

In this section, we shed some light on two essential elements of modern JavaScript code distributions.

\begin{figure}
    \centering
    \includegraphics[width=.87\columnwidth]{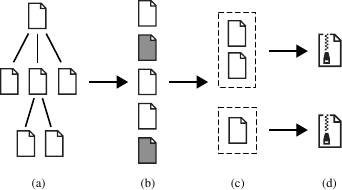}
    \caption{Major phases of the bundling process: (a) module resolution, (b) tree shaking, (c) code splitting, and (d) bundling and minification}
    \label{fig:bundling}
\end{figure}

\subsection{Content Delivery Networks}

\Glspl{cdn} offer hosting services for web resources.
While there are paid providers fulfilling custom requirements, the set of public \glspl{cdn} is of special interest in our context.
These host commonly used JavaScript resources and offer fast loading speeds.
If independent websites include the same resource, it may already be cached by the user's browser.
However, external script inclusions run in the context of the embedding website which raises security concerns.
To solve this, developers can specify cryptographic hashes with the help of subresource integrity~\cite{w3c-sri}.
Then, resources are blocked if the integrity is violated.
Although, some \glspl{cdn} support \emph{version aliasing} where an alias such as \texttt{latest} may be explicitly or implicitly used to reference the most recent matching resource~\cite{jsdelivr-docs}.
These concepts are incompatible and developers must decide what they prioritize.

\subsection{Web Bundling}

To develop ways to reverse engineer JavaScript bundles, it is essential to understand how bundlers work.
According to npm downloads, the most popular bundlers at the time of writing are esbuild, Rollup, Webpack, Browserify and Parcel.
\Cref{fig:bundling} visualizes bundle generation which is split into multiple phases in each of them.

The first phase is called \emph{module resolution}.
Given one or multiple entry points, bundlers resolve all imported dependencies transitively.
This results in a set of selected packages and their dependency graph.
With help of this dependency graph, \emph{tree shaking}, a variant of dead code elimination, may be performed in a second step.
It analyzes the import relationships to discard unused source files.
While this phase is optional for most bundlers, they differ in whether tree shaking is enabled by default.
Utilizing the dynamic import system, the optional phase \emph{code split} is executed.
It separates the module graph on boundaries defined by dynamic imports and produces multiple target bundles.
This reduces the size of the main bundle and, therefore, speeds up the initial page load.
The secondary bundles are loaded on-demand.
The fourth phase is mandatory and involves the real \emph{bundling}.
All modules selected to be placed in the same bundle are arranged as compartments in a module object or array.
Most bundlers support CommonJS (CJS) inclusions through \texttt{require()} and more modern ECMAScript module (ESM) inclusions which utilize the \texttt{import} keyword.
When generating the bundle, both systems are replaced by an access to the module object, referencing the identifier of the compartment corresponding to the original reference.
Some bundlers behave differently in this phase.
esbuild stores each compartment in their own variable while Rollup copies all objects into the global namespace.
Consequently, compartment boundaries and their import relationships cannot be restored for these bundlers later on. 

Last, bundles are minified. 
\emph{Minification} can range from the removal of whitespace over mangling of identifier names to constant propagation and usage of alternative condensed syntax (cf. \Cref{tab:minification}).
This does not alter the behavior of the code, except for unsafe minification procedures which operate more aggressively by utilizing semantic equivalences.
Noticeably, these often require that the prototypes of global objects have not been altered, i.e., well-known functions from the standard library behave as expected.
In order to facilitate the use of existing debugging tools, bundlers have the option to generate source maps containing the sources along with a mapping to the minified artifact.

\begin{table}
    \caption{Examples for minification operations}
    \centering
    \begin{tabular}{lll}
        \toprule
        \textbf{Mechanism} & \textbf{Input} & \textbf{Output} \\\midrule
        \multirow{2}{*}{Constant Propagation} & \verb|"abc" + "def"| & \verb|"abcdef"| \\
        & \verb|24 + 18| & \verb|42| \\\midrule
        \multirow{2}{*}{Unsafe Transformations} & \verb|1+Number(x)+2| & \verb|3+ +x| \\
        & \verb|new Object()| & \verb|{}| \\\midrule
        \multirow{2}{*}{Version Transpilation} & \verb|`a${x}`| & \verb|"a".concat(x)| \\
        & \verb|x ?? 0| & \verb|null!=x?x:0| \\\bottomrule
    \end{tabular}
    \label{tab:minification}
\end{table}

\subsection{Fundamental Approaches}

In this subsection, we describe three existing approaches for package and version detection inside bundles.
While our method does not immediately extend these, their strengths and limitations motivate our fundamental design choices. 
Additionally, they provide a comprehensive view on the space of extractable features.

\citeauthor{Rack2023}~\cite{Rack2023} present an exploratory study of bundles including a methodology for detecting packages and their versions.
Their package detection approach is based on extracting names from source maps.
In contrast, the version detection method is more sophisticated and extracts compartment signatures through string literals, selected keyword usages, and built-in calls.
Each reference package version is downloaded, bundled for preprocessing and its signature indexed in a database.
When extracting the signature from an unknown bundle, a version from the database is detected if at least 80\% of features match.
It should be noted that the authors employ their approach solely for lodash, which includes its own version as a string literal, and do not evaluate its performance.
We will refer to this approach as \emph{BundlerStudy}, the name of their code repository.

\citeauthor{Liu2023}~\cite{Liu2023} introduce \emph{PTdetector}, an approach to detect bundled packages without relying on the presence of metadata.
Their concept of property trees is a dynamic analysis method which creates signatures based on properties of global objects.
They integrate property data types as well as their values if representable.
For objects, this structure is nested.
Their evaluation shows a high performance on utility packages while falling short for frameworks.
Although not explicitly mentioned, PTdetector is conceptually restricted to work on package bundles.
The more general form of application bundles is typically fully self-contained and does not alter global objects.

\citeauthor{Ali2024}~\cite{Ali2024} present \emph{URR} which focuses on detecting and removing selected tracking packages from JavaScript bundles.
As features, URR analyzes the \gls{ast} of each compartment by hashing the token types and their parent-child relationships.
It achieves robustness against some types of minification by discarding all token values.
To build its reference database, the considered tracking packages are bundled with Webpack for all combinations of considered build options.
Consequently, the resulting database takes approximately 1 GB for each reference package.
The evaluation suggests that the performance for \gls{tpl} and version detection is high.
However, as internal parameter values were chosen on the evaluation dataset and their methodology does not offer the possibility to detect false negatives, URR's real-world performance is difficult to be estimated.

\section{JavaScript Bundle Version Detection}
\label{sec:methodology}

With the background knowledge on bundling, we present Aletheia, our approach on extracting version information from bundles.
It requires a bundle and a list of packages as input.
Reliable package identification in bundles is still an open problem.
Similar to prior work~\cite{Rack2023}, we utilize source maps to identify packages through the \verb|node_modules| directory in source file paths.


Depending on the development style of a software project, the difference between two versions may vary greatly.
Generally, the smaller the changesets, the more complex version detection becomes.
Thus, we search for a solution with potential for distinguishing even minor changes.
For arbitrary JavaScript packages, coarse features such as string literals do have potential for package identification~\cite{Rack2023,Ban2021}, but their value diminishes for version detection as they likely stay untouched for patch releases.
Moreover, minification renders text-based comparisons and most identifiers unusable.
Therefore, a successful method for version detection needs to perform of structural or behavioral comparison between a package version and a bundle.

This problem shares traits with source code plagiarism detection.
There, two samples of code are compared to try detecting whether one programmer copied from the other.
Novak et al. correctly identified that proposed solutions are actually similarity detectors as plagiarism cannot be automatically detected~\cite{Novak2019}.
Recently, \emph{Dolos} has been proposed as a source code similarity detector designed for student assignments~\cite{Maertens2022}.
While there are more than a hundred published detection tools~\cite{Novak2019}, Dolos fits our objectives best by being maintained, open-source, well-documented, 
and performs at least on par with other tools~\cite{Maertens2022}.

It works in three phases.
First, Dolos tokenizes the input files into an abstract syntax tree~(AST).
It focuses on structural comparisons by discarding all concrete values and storing only token types, e.g., \verb|Literal|, \verb|Function| or \verb|BinaryExpression|.
Second, it computes a set of fingerprints for each input.
A fingerprint is represented by a hash of $k$ consecutive tokens, a so-called $k$-gram.
To decrease the amount of fingerprints to store and compare, Dolos employs the Winnowing algorithm~\cite{Schleimer2003} with window size $w$.
From each window, it selects the hash with the lowest numerical value.
Last, all fingerprints are indexed, facilitating efficient comparisons.
Two input files are compared by the fingerprints they share in relation to the total amount of fingerprints they contain.
This metric is called \emph{similarity}.

To integrate rolling \gls{ast} hashes to Aletheia, the source files of each package version are combined into a single source file.
As the processing is done independently on a given bundle, the database of known packages can be pre-indexed.
This greatly improves the runtime on a single bundle.
We experimented with different values for $k$ and $w$ on a few versions of \texttt{axios} and achieved the highest relative similarity score with $k=27$ and $w=15$.
Additionally, the default metric for evaluating the similarity is altered to be calculated as the fraction of shared fingerprints contained in the bundle.
This disregards the fraction of shared fingerprints in a package version which can vary greatly between versions.
The reference implementation of Dolos is written in JavaScript and is rather slow with high memory consumption.
For our longitudinal study, we reimplemented the critical processing steps in C++ and discarded unused metadata and calculations in the reporting stage.
To measure the speed-up we implemented a small single-threaded benchmark, which indexes a 1.1 MB JavaScript bundle three times and then reports the statistics for a single pair.
We observe a performance increase factor of 8 for indexing (1.2 s compared to 9.8 s) and of almost 10,000 for reporting (0.011 s compared to 107 s).

It is fundamental for every structural version detection approach to determine a suitable input for the comparison algorithm.
In our context, npm package artifacts conceptually represent the ideal input as they also serve as input for bundlers.
However, the obtainable artifact archives are not immediately suitable and require preprocessing.
The following sections describe three essential steps of Aletheia that are imperative to a good detection performance.

\subsection{File Selection}
\label{sec:file-selection}

Prior research utilizes files from cdnjs~\cite{Liu2023}, which have no clear connection to bundle contents, or fine-tune their preprocessing for a handful of npm packages~\cite{Ali2024,Rack2023}.
To the best of our knowledge, we are the first to propose an algorithm for extracting relevant source code of arbitrary npm release artifacts.
While it initially may sound like a simple problem, its complexity arises from the fact that npm does not enforce any rules on artifacts, except that the \verb|package.json| file contains the most important metadata, i.e., name, version and dependencies.
In particular, common conventions such as referencing an entry point through \verb|"main"| are not consistently adhered to.
In practice, artifacts are highly diverse and contain code and assets which do not affect bundling.
Additionally, the source code is not always JavaScript but may be expressed in a related language such as JSX, TypeScript or CoffeeScript.
We observed many of such cases despite the file extension indicating JavaScript.
For our approaches, it is of high importance that we can separate the source code from other components in the artifact.
Otherwise, central assumptions for similarity comparisons do not hold.

To solve this, we propose a three-step file selection algorithm for Aletheia.
In the beginning, it searches for the fields \texttt{jsdelivr} and \texttt{unpkg} in the main \verb|package.json| file.
If available, these point to a pre-bundled artifact which poses an elegant shortcut and immediately solves the problem.
Otherwise, we search for possible entry points through the fields \texttt{main}, \texttt{module} and \texttt{index}.
In case at least one entrypoint is found, we employ Webpack's module resolution algorithm to recursively collect all imported package-level modules while resolving rules for conditional and sub-path exports.
As a last resort, the algorithm falls back to heuristics based on simple naming conventions.
These select all variants of JavaScript source files in the package, excluding \texttt{test}, \texttt{example} and \texttt{vendor} folders.
A detailed technical description is available as part of the distributed artifact.
It should be noted that our attempts with simpler algorithms drastically affected the results, degrading the version detection performance by factors of 1.5 to 2.

\subsection{Transpilation}
\label{sec:methodology-preprocessing-transpilation}

A closer look at the minification algorithms utilized by bundlers reveals that most optimizations besides identifier mangling alter parts of the AST.
This reduces the amount of shared fingerprints and decreases the version detection quality.
An approach for solving this issue is the transpilation of inputs into an intermediate representation (IR) which uniquely expresses equivalent semantics.
More formally, we would like the following equation to hold for any source code $s$ and any minifier $\min$:
\[ \text{IR}(s) = \text{IR}(\min(s)) \]
Furthermore, two semantically different sources should still have different representations.

To the best of our knowledge and despite its usefulness in code comparisons, there has been no prior work on this topic.
Therefore, we build upon existing tooling and observe that, in fact, minifiers are close to fulfilling the requirements by trying to map sources to their most compact representation.
Consequently, Aletheia uses the subset of JavaScript which is returned by minifiers as intermediate language.
An evaluation of the most common minifiers in an aggressive setting show that swc outperforms other candidates.
It achieves an average similarity score of 0.93 on a representative set of npm packages across multiple scenarios.
This is close to the perfect matching score of 1 of an ideal transpiler.

\subsection{Bundling}

Rolling AST hashes compute the similarity between a pair of files.
Since most packages have their sources distributed over multiple files, we utilize a rudimentary bundling algorithm which wraps all source files into their own function and collects them in an array.
Import and export statements (ESM-style) are rewritten to CJS style which matches the syntax generated by real bundlers more closely.
The resulting bundle is dysfunctional but remains suitable for static analysis.
It serves as the foundation for the similarity calculations with actual bundles.

\section{Evaluation Methodology}
\label{sec:evaluation}

The performance of Aletheia is evaluated in two different scenarios.
First, we evaluate its capabilities on lab-generated bundles.
This has the advantage that we precisely control the ground truth and the environment.
Second, we evaluate Aletheia on bundles originating from Internet measurements, which are more diverse and can be regarded as more challenging.
These contain unique first-party code and potentially reused code not observable in lab bundles.
We compare our performance to BundlerStudy~\cite{Rack2023}.
Initially, we aimed to include URR~\cite{Ali2024} in our analysis as well.
Unfortunately, the repository referenced in its publication does not contain any of the promised material and the authors did not respond to our inquiries.
A re-implementation is hampered as their textual algorithm description lacks details such as the utilized hash function and the conversion of node types into strings.
More severely, it is unclear how the necessary breadth of training data for arbitrary packages with few usages can be obtained in our context while still upholding the split of the validation data.
Therefore, we refrain from an immediate comparison and fall back to a theoretical discussion.
Other approaches~\cite{Pagon2023,Liu2023} are not applicable to the general scenario we consider.

Aletheia compares a bundle to a known reference database which should contain a suitable representation of packages commonly used in the web.
There are two possibilities to obtain appropriate package lists.
First, public CDNs with a fixed set of available packages can be used.
An example for this is cdnjs.
Second, package names extracted from real-world source maps can be evaluated.
We decide to employ the latter, since we can be certain these packages are indeed used inside web bundles and directly obtain npm package names.
Aletheia's precomputed index of all 7,213 packages obtained through this approach has a size of 2.8 GB encoded as JSON.

\begin{figure*}
    \centering
    \includegraphics[width=\textwidth]{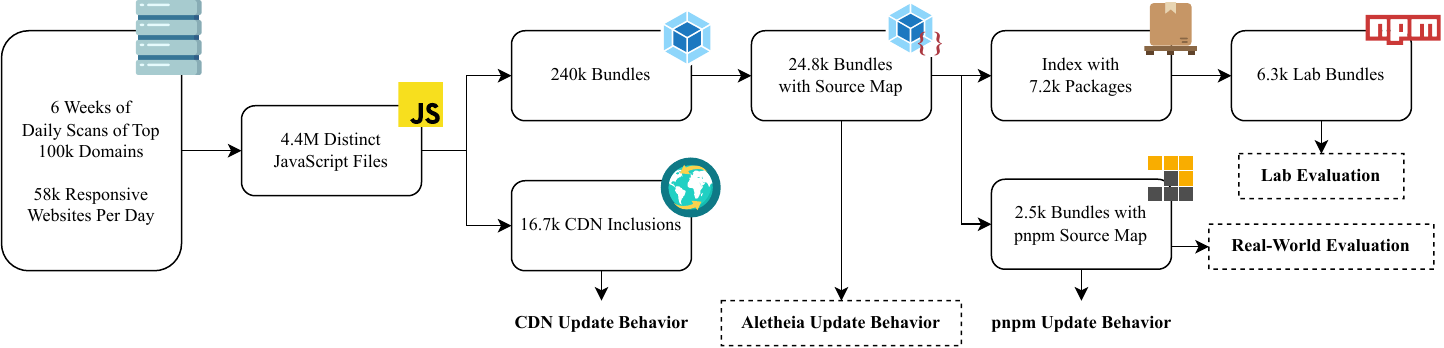}
    \caption{Processing pipelines from the files collected through the Internet scans to the evaluations. The package index is used as the reference database for Aletheia in all evaluations with a dashed border.}
    \label{fig:measurement-flow}
\end{figure*}

It must be noted that not all JavaScript assets available on websites are bundles.
To distinguish JavaScript bundles from normal scripts, bundler fingerprints~\cite{Rack2023} are used.
We improve runtime of this algorithm by flattening the AST with an in-order traversal to a string over the alphabet of token types.
Afterwards, we can use the efficient Aho-Corasick algorithm~\cite{AhoCorasick} to find any matches of any fingerprint within a single pass.
Additionally, after confirming the detection capabilities of the reference fingerprints, we utilize another fingerprint which is able to detect Webpack chunks.
This is a format commonly found on sites with Webpack bundles and generated when code split is activated.

\subsection{Lab-generated Bundles}

For lab-generated bundles, we bundle the latest version of each package including all of its dependencies.
We employ Webpack as bundler with code split disabled.
While other bundlers utilize distinct algorithms, we claim this to be a representative setup.
Compartment extraction works similarly for all bundlers without scope hoisting and differences in minification is accounted for by the transpilation step during preprocessing.
Furthermore, Webpack is by far the most popular bundler found in the wild (see \Cref{sec:bundler-prevalence}).

\subsection{Internet Measurements}

To answer RQ1 and RQ3, we performed daily scans of a Tranco Top 100K domain list\footnote{List available at \url{https://tranco-list.eu/list/G6LKK}}~\cite{LePochat2019} for six weeks starting in November 2024.
The list is filtered to contain only one top-level domain per organization as we expect bundles reuse over different top-level domains.

\subsubsection{Crawler}

Our web crawler uses a headless Chrome instance managed by Puppeteer and honors established methodological guidelines for reproducible Internet measurements~\cite{Demir2022}.
The open-source project \texttt{puppeteer-real-browser} is deployed for bot detection evasion.
While it is unmaintained at the time of writing, it has been actively maintained during the measurements.
This aspect hinders future reproduction of our measurements but is a typical challenge encountered in Internet measurement research~\cite{Jueckstock2021,Demir2022}.
Furthermore, the domain list is shuffled each day and the crawler cycles through six different source IPv4 addresses.
To capture dynamically loaded scripts, every page is kept open for 15 seconds.
Past research has shown that longer timeouts have negligible effects on the results~\cite{Jueckstock2021}.
For each page load, we record all requested scripts and store them into the dataset.
In case they reference external source maps through comments or HTTP headers, we try to fetch them immediately.
Inline source maps do not require additional requests.

For every domain, we request the landing page via a TLS connection.
This reduces our hit rate but allows us to exclude websites which are not maintained and, therefore, should not be regarded as production setups.
Overall, the scanning parameters are chosen conservatively and discussed in \Cref{sec:ethical-considerations}.

\subsubsection{Ground Truth}
\label{sec:ground-truth}

The major issue with an evaluation on real JavaScript bundles is to obtain a suitable ground truth for the analysis.
We find source maps to be the only additional source of information which can be utilized automatically.
Though it must be noted, that they are not available for all bundles.

Similar to the package selection described before, we utilize the source file paths from the source map to derive the packages contained in a bundle.
For the special case that the developer used \texttt{pnpm} instead of \texttt{npm} as package manager, we can even obtain the ground truth versions of the packages:
\texttt{pnpm} de-duplicates packages by symlinking to a central store containing all downloaded packages.
As it also supports monorepos, i.e., single repositories with multiple projects, it must be able to manage multiple versions of the same package.
For this, it integrates the version number into the directory name.
Webpack resolves these symlinks and stores the full path including the version of a package in the source map.



\subsection{Metrics}

This section presents metrics for answering RQ2 and RQ3.

\subsubsection{Version Detection}

Version detection results in the output of one or multiple versions.
We decide to employ two different metrics capturing distinct aspects.
The first metric is called \emph{version difference}.
We interpret versions as vectors and calculate their difference.
If the detection returns multiple versions, we use the smallest difference.
For simplicity, prerelease specifiers such as \texttt{1.2.3-rc1} are mapped to the corresponding core version (\texttt{1.2.3}) while early development versions do not receive special treatment.
We argue that this decision does not negatively impact the metric since, while npm strongly recommends the use of the Semantic Versioning scheme~\cite{npm2024semantic}, there are indicators that a significant share does not strictly follow the associated rules~\cite{Dietrich2019}.
Furthermore, different update strategies range from small updates every day to large updates every year~\cite{Pinckney2023}.
This distorts the results as the meaning of version differences needs to be interpreted differently for each project.
The second metric \emph{difference existence} improves upon that.
It reduces the complexity of the version difference by replacing it with the triple binary output, whether a major/minor/patch-level difference exists between the correct and the detected version or not.
For ranged results, we follow the strategy from the first metric.
To prevent coincidental performance fluctuations, differences in the major version always imply an erroneous detection of the minor- and patch-level, and analogously do minor component errors imply patch-level errors.


\subsubsection{Update Patterns}

To characterize general update patterns of packages, we analyze the \emph{rollout time} for updates.
The rollout time starts with the release date of a new package version and ends with the first measurement in which the new version is observed.
For our dataset, we analyze the fraction of domains which achieve certain rollout times.
We decided to loosely associate our thresholds with different scenarios.
Updates within one week are an indicator for an automated update procedure.
Four weeks are the median interval between CVE assignment and vulnerability disclosure~\cite{Li2017} and may be regarded as a threshold for security-aware developers.
As the longest time frame we choose 16 weeks as an interval of general active software maintenance.
In addition, we determine the \emph{prevalence of vulnerabilities} by connecting the extracted version information with advisories from the Snyk Vulnerability Database\footnote{\url{https://security.snyk.io}}.
This metric is measured by the mean amount of vulnerable packages loaded by a domain.
Though, this metric must be taken with care, since there might be multiple prerequisites to be fulfilled so that a vulnerability propagates into the web application and is exploitable.

\section{Results}
\label{sec:results}

For crawling, a server with an Intel Xeon E5-4640 v4 (96 logical cores) CPU and 512 GB of RAM was employed.
Afterwards, we evaluate our datasets on an HPC compute node (96 cores, 1 TB RAM) since most of our computations can be parallelized.
\Cref{fig:measurement-flow} visualizes the dataset sizes at different stages of the processing pipeline.

\subsection{JavaScript on the Web}

In this section, we analyze the prevalence of JavaScript bundles and CDN resource inclusions on the web.
This allows us to examine recent developments in this area by comparing the results with previous studies.

\subsubsection{Bundler Prevalence}
\label{sec:bundler-prevalence}

Our dataset contains bundles matching fingerprints of multiple bundlers.
Manual inspection reveals unanticipated cases of bundles inside other bundles.
In one particular case, a Webpack chunk contains a single compartment with a bundle of the npm package \texttt{dc-delivery-sdk-js} inside.  
Closer examination shows that its release artifact contains all of its source code files as well as bundles for different target platforms.
During bundling, the \texttt{browser} field of the \texttt{package.json} is the decisive factor for Webpack to select the pre-bundled sources of the other files.
It should be noted that the artifact does contain a source map for the bundle, but it is inevitably lost during the second bundling as no bundler supports the merging of bundles and their source maps at the time of writing.
For this analysis, we discard bundles matching multiple fingerprints as we cannot attribute the bundler reliably.

\begin{figure}
    \centering
    \includegraphics[width=\columnwidth]{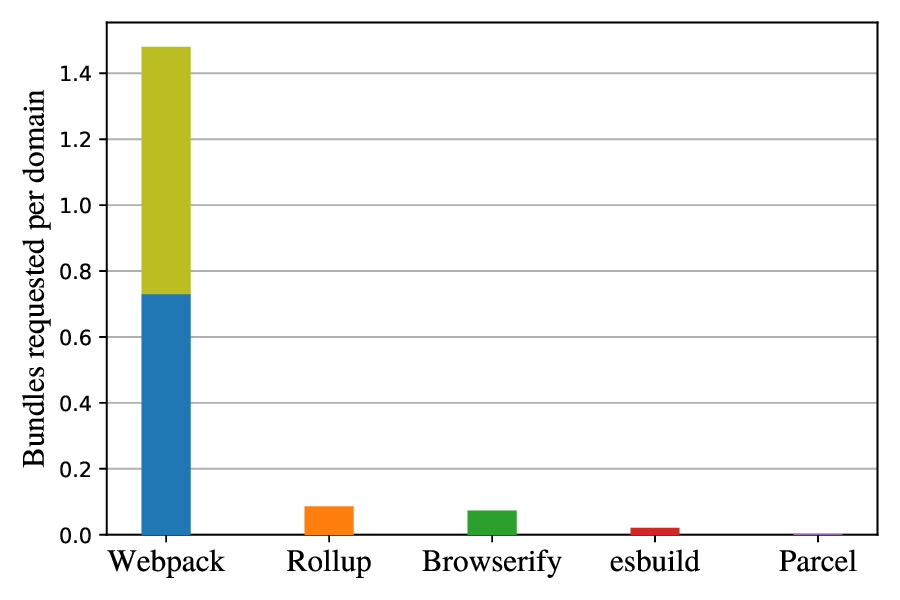}
    \caption{Average amounts of bundles per domain and bundler. Webpack chunks (yellow) are responsible for half of its detections.}
    \label{fig:prevalence-bundler}
\end{figure}

The results are shown in \Cref{fig:prevalence-bundler}.
Overall, we find 1.45 (Webpack, including chunks), 0.09 (Rollup), 0.07 (Browserify), 0.02 (esbuild), and 0.002 (Parcel) mean bundles included per domain which largely replicates previous results~\cite{Rack2023}.
Therefore, we conclude that \textbf{bundles remain highly prevalent throughout the top 100k websites} (RQ1).
Observable developments are Rollup and esbuild with double the average amount of bundles.
This is explained by an evolving JavaScript ecosystem between the two measurements:
npm download statistics clearly indicate an increase of popularity for esbuild and Rollup but not for the other bundlers.

\begin{figure}
    \centering
    \includegraphics[width=\columnwidth]{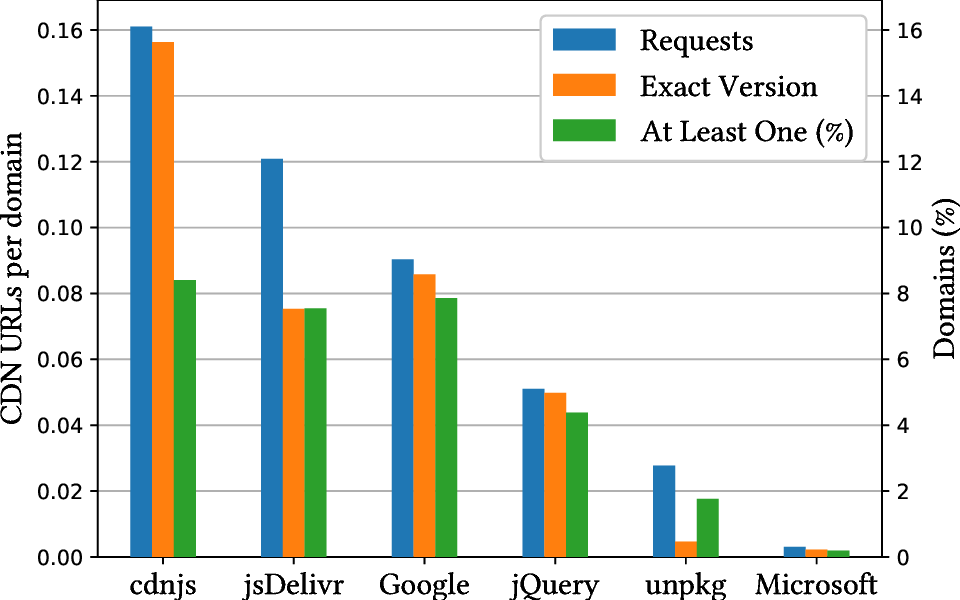}
    \caption{Average absolute amounts of included CDN \glspl{url} per responsive domain and CDN. \emph{At least one} denotes the percentage of domains with a CDN resource.}
    \label{fig:cdn-urls}
\end{figure}

\subsubsection{CDN Usage}
\label{sec:cdn-usage}

We analyze script inclusions from the open source CDNs cdnjs, jsDelivr, from commercial CDNs hosted by Google and Microsoft, and from jQuery's library-branded CDN.
In addition, we analyze \glspl{url} from unpkg, another popular open-source CDN serving any npm package similar to jsDelivr.
In contrast to previous work~\cite{Lauinger2017}, we refrain from including Bootstrap CDN as it has been merged into jsDelivr in 2021 and only few \glspl{url} appeared during our web scans.

\Cref{fig:cdn-urls} visualizes the results.
We find on average every other domain to include a CDN asset.
cdnjs and jsDelivr have the highest prevalence followed by Google and jQuery.
As the latter two only serve few packages, most domains include only a single resource from these.
Microsoft's CDN is used very sparsely and can thus be regarded as negligible.
It is observable that most domains query a concrete patch-level package version for cdnjs, Google and jQuery.
In contrast, every other jsDelivr and unpkg \gls{url} refers to version-aliased artifacts which are automatically updated.
This result is important when considering update patterns in \Cref{sec:update-behavior}.
Overall, \textbf{CDN \glspl{url} have a significantly lower prevalence compared to bundles in our dataset} (RQ1).
Though, it must be noted that CDN artifacts may in fact be bundles themselves.

\subsection{File Selection Algorithm}
\label{sec:file-selection-results}

Aletheia's file selection algorithm presented in \Cref{sec:file-selection} is an essential component for the package and version detection algorithms.
We evaluate its performance by selecting 100 package versions and comparing it with our manual assessment.
For this, we first sample packages in a uniformly random manner and then choose a version analogously.
This prevents a bias towards packages with a high amount of versions.
The manual assessment is mapped to one of three categories: perfect, equivalent (semantically identical, e.g., source files instead of minified artifact), and erroneous.
Overall, \textbf{we found the selection to be perfect or equivalent in 97\% of cases} (RQ2).
Due to the uniform sample, the confidence interval for sample proportions of Bernoulli variables can be calculated.
Then, the performance of the file selection algorithm is estimated to lie within the interval $[0.937,1]$ with a probability of 95\%.

\subsection{Version Detection on Lab Bundles}


The following results are based on 6,337 generated lab bundles.
These are 88\% of packages which are used to build Aletheia's index.
For the remainder, automatic bundling with Webpack failed, most commonly due to the absence of an entry point.

\Cref{tab:results} provides an overview over the results.
It must be noted that each bundle contains a main package as well as all of its dependencies.
Therefore, the difference existence metric differs whether grouped by packages or not.
Overall, \textbf{Aletheia is able to detect the correct patch levels in 87\% of instances and 87.2\% of packages in the lab setting} (RQ2).
Even without compartment information, it is able to achieve detection rates of 81.8\% and 81\%, respectively.
The mean version component error is below one in most cases.
In contrast, even with compartment information, BundlerStudy clearly performs worse in all metrics, achieving patch-level detection rates of 48.3\% and 59.8\%.
In this case, the mean error needs to be interpreted carefully:
One of the wrongly classified packages utilizes date-based patch identifiers which leads to a case with the patch-level error amounting to 202405101233.
Such a date-based patch number strongly affects the resulting mean error and highlights the limitations of this metric.

\begin{table*}
    \centering
    \caption{Quantitative results when comparing Aletheia to BundlerStudy. \emph{Correct} refers to the difference existence metric. Combined statistics correspond to values when considering major/minor/patch versions. The performance of the algorithms without compartment information is denoted by \textdagger.}
    \begin{tabular}{l|ccc|ccc}
        \toprule
                                       & \multicolumn{3}{c}{\textbf{Lab Setting}} & \multicolumn{3}{|c}{\textbf{Real-World Setting}} \\
                                       & Correct & Correct by Pkg & Mean Error & Correct & Correct by Pkg & Mean Error \\ \midrule
         Aletheia                         & 97.3\%/90.1\%/\textbf{87.0\%} & 95.8\%/90.5\%/\textbf{87.2\%} & 0.05/0.64/\textbf{0.26} & 93.9\%/86.7\%/\textbf{82.1\%} & 92.6\%/82.8\%/\textbf{77.0\%} & 0.10/1.81/\textbf{0.45} \\
         Aletheia\textsuperscript\textdagger               & 94.1\%/85.4\%/\textbf{81.8\%} & 92.0\%/84.9\%/\textbf{81.0\%} & 0.20/1.06/\textbf{0.40} & 92.3\%/82.1\%/\textbf{77.1\%} & 89.7\%/76.7\%/\textbf{70.4\%} & 0.15/2.02/\textbf{0.76} \\
         BundlerStudy                  & 78.8\%/54.9\%/\textbf{48.3\%} & 83.5\%/67.3\%/\textbf{59.8\%} & 0.47/2.97/\textbf{8$\cdot$10\textsuperscript 6} & 73.9\%/48.3\%/\textbf{41.7\%} & 74.3\%/50.4\%/\textbf{43.2\%} & 0.62/5.54/\textbf{2.39} \\
         BundlerStudy\textsuperscript\textdagger        & 68.4\%/47.8\%/\textbf{39.9\%} & 69.9\%/52.2\%/\textbf{45.0\%} & 1.18/679/\textbf{6$\cdot$10\textsuperscript 6} & 60.8\%/29.9\%/\textbf{22.3\%} & 58.7\%/32.5\%/\textbf{24.6\%} & 1.20/5.57/\textbf{3.14}\\\bottomrule
    \end{tabular}
    \label{tab:results}
\end{table*}

\subsection{Version Detection on Real-world Bundles}

Our measurement dataset comprises 24,758 distinct bundles with a suitable source map as ground truth which we use to extend our answer to RQ2.
A subset of 2,516 bundles has pnpm-generated source maps functioning as ground truth for version detection.
It must be noted that Aletheia's index is built based on package names extracted from this ground truth.
However, this does not cause an overlap of the training and testing sets, since the evaluation focuses on version detection performance and not package identification.

Aletheia could be applied to 4,399 packages cumulated across all suitable bundles.
While each bundle contains much more than two packages on average, many of them are not indexed, caused by a significant share of private packages.
The performance decreases slightly compared to the lab analysis \textbf{with a correct version detection in 82\% of cases (77\% of packages)} (RQ2).
Again, BundlerStudy is not competitive with patch detection rates of 41.7\% (43.2\% of packages).
The detection rates of its compartment-less variant are below 25\%, indicating a strong impact of first party code.
Conversely, the distance between correct detections between both Aletheia variants is close to constant between the scenarios.
This suggests a high robustness of our version detection algorithm when confronted with first party code which also covers potential code reuse.

We deeply inspect a sample of 10 bundles with erroneous classifications to understand causes of Aletheia's misclassifications.
In seven cases, we observe small changesets between the detected and the correct package versions, independent of the version difference which sometimes amounts to several major versions.
Additionally, tree shaking is responsible that in three of these cases the changesets are not even included in the bundle.
Throughout all seven cases, we find noise such as different ways of referencing modules to be responsible for slightly lower similarities of the correct version.
In two cases, an erroneous file selection causes issues.
For example, for \texttt{@babel/runtime-corejs3} 7.10.3 an \texttt{index.js} file in a subdirectory is selected due to a lack of a specified entry point.
As it contains only a single statement, hundreds of relevant source files are ignored.
In the last case, we unexpectedly observe two different versions of the same package inside the bundle.
Aletheia identifies a correct version, just not the one which our ground truth expects.

\subsection{Update Patterns}
\label{sec:update-behavior}

In this section, we perform a longitudinal study to analyze the update patterns and corresponding vulnerability timeframes of JavaScript packages from different sources on the web.
We consider three data partitions: CDN resources, bundles with source maps and pnpm-based bundles with source maps.
As the partitions overlap, we assign their intersections for this analysis as follows.
Bundles served on a CDN \gls{url} are assigned to CDN resources as these are package bundles with an update procedure distinct from application bundles.
pnpm-based bundles are considered separately.
This is motivated by differences in the detection reliability and the partition size.
The versions in pnpm-based bundles can be detected reliably but only few bundles fall into this category.
Conversely, there are many bundles with source maps but Aletheia does not exhibit perfect precision.
To maximize its reliability, we refrain from considering packages with versions that lacked accurate detections in the lab or real-world scenarios.

\begin{figure}
    \centering
    \includegraphics[width=\columnwidth]{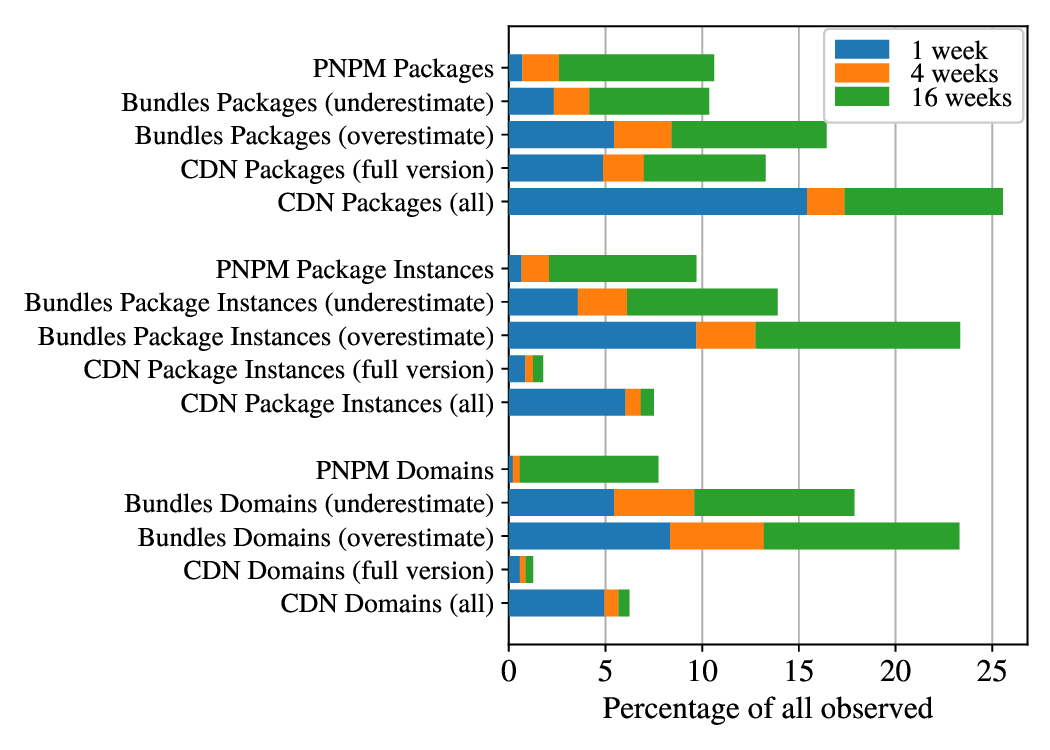}
    \caption{Percentage of packages, package instances and domains where updates onto a version published in the given interval have been observed. Stacked bars are cumulative.}
    \label{fig:update-behavior}
\end{figure}

\Cref{fig:update-behavior} visualizes the results.
For a given rollout time, three categories are distinguished: the fraction of packages and the fraction of package instances, i.e., packages weighted by their prevalence, being updated in time as well as the fraction of domains updating their JavaScript packages in time.
Furthermore, we separate CDN \glspl{url} with a hardcoded version to not mix distinct update procedures.
For bundles, we discard all detection ranges which include more than three versions.
For the remaining data, the figure splits underestimated and overestimated versions, i.e., selecting either the lowest or highest version from the returned range.
We utilize the Wilcoxon-Rank-Sum test from inferential statistics to estimate the confidence into our results.
The tests are calculated on the domain-oriented dataset partition for 16 weeks.

For CDN \glspl{url}, a clear difference between fixed-versioned and all \glspl{url} can be observed.
Considering the results for package instances and domains, the statistic for packages does not seem to resemble the breadth of the web as well since niche packages are represented too strongly.
Therefore, RQ3 is answered with \textbf{version aliasing significantly improves the rollout time of dependencies included in CDN \glspl{url} ($U=10.1,~p<0.01$) but they still show a worse update adoption than bundles after 16 weeks} ($U=14.7,~p<0.01$ for underestimation).
Though notably, most updates occur within the first week, rendering version aliasing at least on par in that regard.
For bundles, the pnpm partition unexpectedly exhibits vastly distinct update patterns compared to general bundles.
\textbf{Only 0.6\% of domains utilizing pnpm update their packages within 4 weeks, while in the following 12 weeks this percentage rises to 7.7\%, surpassing CDN \glspl{url}} ($U=5.34,~p<0.01$) (RQ3).
In contrast, general bundles see smoother distributions between the analyzed time intervals.
Additionally, \textbf{more than 18\% of domains update their bundles within 16 weeks} (RQ3).
Even when taking possible misdetections into account, this is a significant margin over all other partitions and indicates a positive development.
Still, this implies more than 76\% of domains do not update their bundles within 16 weeks, allowing attackers to possibly exploit known vulnerabilities for an extended period of time.

For the general prevalence of vulnerabilities, we observe 35\% of CDN-included packages per domain to contain a vulnerability. 
By two orders of magnitude, \texttt{jquery} has the most vulnerable package instances.
In contrast, on average only 5\% of packages included in bundles with pnpm source maps have known vulnerabilities.
Half of the occurrences correspond to old versions of \texttt{webpack}.
Despite being a bundler, weaknesses in its runtime or the asset generation such as CVE-2024-43788 may impact the security of the final web application.
Through Aletheia, we find a comparable mean of 2.7\% -- 3.7\% of vulnerable package versions included in bundles per domain, depending on the type of estimation.
Most commonly, we find ReDoS vulnerabilities in \texttt{vue} (CVE-2024-9506) and \texttt{moment} (CVE-2022-31129).
Overall, it is notable that \textbf{CDN \glspl{url} are 7x -- 10x more likely to include a vulnerable package compared to bundles}.
Considering the slow update behavior, this poses a significant security threat.


\section{Discussion}
\label{sec:discussion}

In this section, we discuss our methodology and results while considering limitations and future work.

\subsubsection*{JavaScript package prevalence}


Compared to previous studies, our experimental setup differs in the toplist, crawler, hit rate, and bot detection evasion techniques used.
Thus, the analysis goes beyond reproduction and can be classified as replication.
In general, replication of Internet measurements is complicated as many small factors may have a large impact~\cite{Jueckstock2021,Demir2022}.
Consequently, it is notable that our results for bundlers are in line with the results from \citeauthor{Rack2023}~\cite{Rack2023}.
The few observed differences in bundler prevalence are well-explainable with the evolution of the JavaScript ecosystem and align with corresponding package popularity measures.

For CDN resources, a comparison of our results with measurements conducted by \citeauthor{Lauinger2017}~\cite{Lauinger2017} in May 2016 reveals a shift in popularity.
Back then, 17.6\% of Alexa top 75,000 websites included resources from Google's CDN, followed by jQuery (3.5\%) and cdnjs (2.1\%).
jsDelivr and unpkg were not established at that point in time.
All CDNs with increased popularity share the property that every package listed on npm is accessible.
In comparison, Google's CDN only offers 20 selected packages.
The youngest package has been created in 2018 while the median age amounts to 13 years.
Additionally, most packages are deprecated and unmaintained which explains its decline.
Interestingly, jQuery's CDN is able to maintain its popularity.
As 25\% of observed versions requested from this CDN are released in 2016 or earlier, a possible explanation for its prevalence is that websites which use jQuery are rather old and unmaintained but continue to exist.
This argument is enforced as many of its functionalities are covered by modern browser APIs.
Overall, these developments indicate that the JavaScript ecosystem is more centered around npm compared to 2016.

\subsubsection*{File selection algorithm}


The performance of the file selection algorithm is deemed to be good but not great.
For Aletheia, even few cases of a wrong or an inconsistent file selection impact its performance as seen for the manually analyzed samples.
Despite, we deem the potential for further improvements of the algorithm to be low.
Packages on npm are too diverse and the rules enforced on release artifacts from npm's early days are too limited.
Instead, we anticipate performance improvements by developing an anomaly detection which is able to identify and exclude package versions with erroneous selections.
Hereby, the effects of a wrong selection can be mitigated.

\subsubsection*{Version detection}

Source code offers lots of features for structural analysis.
Existing version detection techniques focus either on standalone packages or are handcrafted detection mechanisms for few popular packages.
While these work well in their specific use case, neither of them works package-agnostic and is able to analyze application bundles.
This naturally confines their utility for analyzing updates of web application supply chains.
The release cycles for popular packages might differ from niche packages while at the same time package bundles require an inherent different update procedure than other JavaScript resources.

In Aletheia, we employ static structural analysis to compare a bundle with known package versions.
Plagiarism detection algorithms excel in this application context and are able to spot even minor differences.
Furthermore, they do not rely on exact matches of the AST structure such as URR~\cite{Ali2024} which requires a database of all possible module transformations.
The results confirm a good performance but also reveal a gap to a perfect detection.
Based on the manual analysis, we identify two primary root causes, namely automatic code transformations executed during bundling and erroneous sets of files associated with a package version.
The real-world setting causes a slightly stronger drop in performance when utilizing compartment information than without.
This is explained by the increased diversity of the real-world bundles affecting the clean compartment-based lab scenario more than the compartment-less execution where rolling hash is already impacted by imperfect inputs.
On the positive side, this result implies that we could not observe a strong impact of website-specific first party code and code reuse on Aletheia.
Instead, the most severe effect is caused by code transformations during bundling.
As an example, during development, we observed detections of all \texttt{jquery} instances as version 1.9.1 regardless of their real version.
Closer inspection revealed that this specific version contained a minified artifact which was not present for all other versions.
While swc was able to solve this specific case, a custom-tailored transpiler would likely improve Aletheia's performance and, additionally, be applicable in other contexts.
For example, it could increase the robustness of automatic detection of malware patterns on npm~\cite{doll2019automated,ohm2022towards,microsoft-ossgadget-backdoor}.

Our core similarity comparison algorithm, the rolling AST hash of Dolos, offers further optimization potential as well.
A central aspect of it involves focusing solely on the types of AST nodes.
While this may be ideal to robustly find similarities between student assignments, it unnecessarily discards some information usable for version detection.
Minification alters the majority of tokens but keeps the value of many constants as well as the types of most operators.
By incorporating this information into the rolling hash, the likelihood of collisions with unrelated could be decreased.

Despite this optimization potential, version detection based on similarity detection is conceptually limited in some cases.
First, heavy tree shaking used for utility packages such as \texttt{core-js} 
may discard most parts of the code base.
Especially, if modules and functions are relatively short and reordered, the detection performance is reduced.
Next, packages offering special assets like icon sets might not offer any source code which changes between versions.
However, in this case updates are not pertinent to security and of limited interest.
Last, micro-packages may not offer enough fingerprinting surface.
Thus, boilerplate code such as a bundler runtime impact the similarity measures above average.
To reduce the impact of such noise, the threshold for the similarity score of detected versions could be lowered, for example, to 90\% of the largest observed similarity.
This would have reduced most incorrect detections in our manually analyzed samples and likely increases the general precision at the cost of a lower recall.

Our reimplementation of Dolos greatly improves the resource consumption in the given use case.
Aletheia's index size of 2.8 GB for 7,213 packages may seem large but is a clear improvement to a database size of 2.3 GB to 3.6 GB for the three packages analyzed by URR~\cite{Ali2024}.
While it is recommended that bundle chunks have sizes of at most a few hundred kilobytes\footnote{\url{https://vite.dev/config/build-options.html\#build-chunksizewarninglimit}}, we found plenty of bundles with sizes of multiple megabytes.
For these, Aletheia needs to hash millions of extracted $k$-grams.
This currently limits the potential for analyzing bundles on-the-fly during a page load.


\subsubsection*{Internet measurements and update patterns}

Analysis of package update patterns through Internet measurements crucially assumes there is no adversary that deliberately obfuscates their bundles in order to hide version critical information.
Despite not regarded as worthwhile, security by obscurity is employed in some areas of the web.
An example is WordPress where plugins exist that remove the version from all outputs~\cite{sucuri}.
However, for JavaScript packages there are neither automatic attack scenarios nor known version detection mechanisms and known attacks utilizing version information from bundles.
Therefore, we neither see an incentive nor a simple way for administrators to identify and modify critical parts of their bundles.
In contrast, it is known that Internet measurements are affected by bot detection which threaten their validity~\cite{Jueckstock2021}.
Unfortunately, there is no up-to-date tool for exposing bot detectors with past work reasoning that bot detection mechanisms are constantly evolving and so need their revelation techniques~\cite{Jonker2019}.
We argue that the successful replication described in \Cref{sec:bundler-prevalence} implies an average impact of bot detection on our measurements.

Overall, the results indicate that the majority of web applications does not exhibit active update patterns.
While similar results were shown multiple times in the past~\cite{Lauinger2017,Pagon2023,Stock2017}, our results reveal two positive trends.
On the one hand, automatic updates through CDNs significantly improve the situation compared to the fixed-version inclusions available in the past.
Priorly, only a small fraction of websites utilized version aliasing and it has been portrayed rather negatively~\cite{Lauinger2017}.
On the other hand, JavaScript bundling as a more modern development style seems to positively impact update patterns.
Both could be simply explained by software updates being viewed as a key security advice for more than a decade~\cite{Redmiles2016} and this belief seems to slowly but steadily permeate developers' actions.
We argue that a more differentiated discussion is necessary.

First, pnpm as a tool is younger than the concept of bundling but exhibits worse update patterns, especially in the first four weeks.
Its main advantage over the default npm is the deduplication of dependencies into a central storage and the compatibility provided through symbolic links.
This is able to save disk space for the \verb|node_modules| directory, which is beneficial on developer's PCs where lots of dependencies tend to accumulate over time.
In contrast, updates in the first week likely correspond to automated setups based on continuous integration and delivery (CI/CD) pipelines.
Usually, every CI run uses a fresh container.
pnpm is not part of the default NodeJS distribution and requires an additional installation step.
Therefore, the update patterns seen for pnpm primarily correspond to manual deployments explaining the observed delays.

Second, we note a significant effect by large vendors onto the detected updates.
Exemplary, we examine Shopify, a prominent vendor offering solutions and components to develop online shops.
During examination of the CDN inclusions, we noticed the amount of updates of fixed versions to be higher than expected with regard to the previous literature.
We find two Shopify extension packages, \texttt{@letscooee/web-sdk} and \texttt{@appmate/wishlist}, to be responsible for more than 40\% of fixed-version CDN \gls{url} updates within the top 100,000 domains.
Notably, Shopify has adjusted its package distribution approach at the time of writing, and both packages are no longer requested from a CDN.
It is likely that Shopify has a measurable effect on the bundle partitions as well.
Other widespread services, such as OneTrust's cookie consent management which is detectable through \texttt{cdn.cookielaw.org} \glspl{url} and prevalent on 5\% of our requested domains, can have comparably strong effects.
At first, 5\% may appear as a small fraction of domains.
Though, when taking into account that, regardless of the partition, no more than 15\% of domains updated their resources within four weeks, such services can have a significant impact on the statistics.

From a deductive point of view, the simplicity of including version-aliased CDN resources has clear usability advantages over the complexity of updating JavaScript bundles.
Usually, when initializing a new project with a popular frontend framework, a pipeline of multiple preprocessing steps such as linters, compilers and bundlers is set up.
In this scenario, updating a production environment requires updating dependencies by running the package manager, running the pipeline to build an artifact and copying the artifact onto a server.
These steps can be automated with dependency management bots and CI/CD pipelines but this requires additional expertise to be set up.
In addition, it often requires a chain of developers, operators and administrative users to install an update in a production environment which fails if a single link slacks off.
Future research should try to separate these large vendor realms to provide further insights into the causal backgrounds of our observations.

Aletheia requires source maps to be present for bundles and is, similar to prior research~\cite{Rack2023}, based on the assumption that this subset is representative for all bundles.
Concerns can be expressed as source maps on production servers may seem like a bad security practice, which could imply non-representative update patterns.
We argue conversely, that there are no known indications that an available source map allows to infer security levels of a web application.
In fact, their absence could also be interpreted as following security by obscurity or utilizing outdated technologies not capable of source map generation.
Still, it is necessary to analyze and question the validity of this assumption in future research.

It should be noted that the web is constantly developing and the applicability of our methodology may change.
As an example, the npm steering committee recently decided to stop distributing Corepack together with NodeJS~\cite{corepack-distribution}.
This may have a negative impact on the usage of alternative package managers including pnpm.
Consequently, the limited observations of pnpm-based source maps may experience a further decline in the future.
Similarly, the prevalence of source maps on production servers is subject to change.
Therefore, we advocate to develop package identification techniques for bundles which go beyond the current state of the art~\cite{Liu2023} and work reliably in the general case.

Furthermore, there are aspects extending beyond the main scope of this paper that influence observed update patterns of client-side JavaScript packages.
Version specifiers may deliberately block updates, e.g., if the major version changes, to prevent incompatibilities to arise.
Across JavaScript packages, it has been shown that every third package release has a pinned dependency that suffers from increasing technical lag~\cite{Decan2018}, which propagates into main applications and is not easily solvable by application developers.
Thus, a timely update of all transitive dependencies is often not realistic.

\section{Related Work}
\label{sec:related-work}

Most closely, this paper advances the area of longitudinal Internet studies in the context of JavaScript.
However, software engineering and security communities have also dedicated considerable research efforts into the closely related fields of \gls{sca} and \gls{tpl} detection.
In this section, we compare existing methods and challenges in these areas to our work.

\subsection{Longitudinal Internet Studies}

\citeauthor{Lauinger2017}~\cite{Lauinger2017} employ a mixture of hash-based detection and dynamic analysis through probing for the existence of known global objects.
They find more than 37\% of domains using at least one vulnerable package.
\citeauthor{Stock2017}~\cite{Stock2017} confirm their results by utilizing Retire.js, a tool with hand-crafted package-specific detection patterns.
Wappalyzer works in a similar way and is utilized by multiple authors~\cite{Demir2021,Lim2023} to analyze the update frequency of the web software stack.
They observe that many websites exhibit potentially insecure update patterns, with some sites updating components too infrequently or running software with known vulnerabilities.
\citeauthor{Pagon2023}~\cite{Pagon2023} detect JavaScript package versions using hash values, identifiers, and differences in source code to similar versions.
Their method is able to detect versions in 30\% of cases.

\subsection{C/C++ Software Composition Analysis}

Motivated by the search of license violations and vulnerability propagation~\cite{Jiang2023}, \gls{sca} is a highly active field of research in the C/C++ ecosystem.
Compared to JavaScript, there exist some specialties such as the lack of a well-established way of distributing dependencies~\cite{Tang2022a}.
Additionally, it is more difficult to obtain a set of annotated binaries from varying build machines.
This manifests in \emph{real-world evaluations} in this realm to correspond to our lab setting as researchers typically build the evaluation binaries themselves.
This is challenging for representative evaluations as indicated by performance drops on different datasets~\cite{Jiang2023,Li2023}.
Other special considerations specific to C/C++ include the effects of function inlining~\cite{Jia2023}, identification of version strings~\cite{Cheng2023} and discarding simple functions~\cite{Wu2023}.
The latter has no relevance for JavaScript as micro-packages may not offer any other features for classification.
However, C/C++ \gls{tpl} detection also shares similar approaches such as obligatory file selection~\cite{Ban2021}, a focus on fused binaries~\cite{Yang2022} which share traits with bundles, and preprocessing as a performance boost~\cite{Wong2024}.
\citeauthor{Wang2024}~\cite{Wang2024} share our fundamental idea and transfer algorithms from the more established area of binary similarity analysis to \gls{sca}.
The compartment dependency graph of bundles could allow methods like the graph-based dependency recall~\cite{Jiang2023} and anchor alignment~\cite{Li2023} to be applied.
Moreover, transformer-based approaches~\cite{Jiang2024,Wang2024a,Zhang2025} also offer distinct advantages and challenges in this context.




\subsection{Android TPL Detection}

\Gls{tpl} detection for Android apps is as well researched as for C/C++ but distinctly motivated by malicious~\cite{Liu2023}, vulnerable~\cite{Zhan2022} and modified \glspl{tpl}~\cite{Wang2018}.
Therefore, the proposed detection methods are typically centered around obfuscation~\cite{Li2017a,Wang2018,Zhang2019,Liu2023} and optimization~\cite{Xie2024} techniques.
The observed techniques share similarities with JavaScript minification~\cite{Moog2021}, despite different intentions.
Consequently, Android \gls{tpl} detection methods employ identifier-independent methods~\cite{Li2017a}, rolling hash algorithms~\cite{Wang2018} and basic block signatures using AST token types~\cite{Zhang2019} which Aletheia uses as well.
Advanced scenarios like runtime code decryption~\cite{Liu2023} are not relevant for typical JavaScript bundles.
In a systematic literature review, \citeauthor{Zhan2022}~\cite{Zhan2022} have shown state-of-the-art detection processes to be comprised of preprocessing, library instance creation, feature extraction and library identification phases which applies to our proposed method as well.
Additionally, they observe version identification to be more challenging than library detection due to the fluctuation of changeset sizes which also holds for JavaScript packages.
Central issues in this field are poor detection performances on different datasets~\cite{Gu2025,Xie2024,Zhan2025}, partially caused by invalid assumptions~\cite{Zhan2025}.
In this regard, the JavaScript bundle analysis field is too young and unexplored yet to perform similar studies.





\section{Ethical Considerations}
\label{sec:ethical-considerations}

When performing Internet scans it is necessary to ethically assess the impact.
Our scans have a low impact by design as they are performed only once per day.
Known malware domains are filtered from the toplist through Google's Safe Browsing database to prevent false alarms in network monitoring systems.
Once started, the scans are throttled to not exceed 8 domain queries per second to limit the stress on networking infrastructure.
For every domain, we request only the landing page which already can generate considerable traffic.
Thus, we block all media requests and refrain from retrying requests if they fail.
Circumventing bot detection is considered as necessary to be able to study the Internet and therefore regarded as ethically unproblematic~\cite{Demir2022}.
We decided not to disclose vulnerable packages detections to web administrators since determining whether a web application is really exploitable would require significant time effort from both parties.

\section{Conclusion}
\label{sec:conclusion}

In this paper, we present Aletheia, a method to detect versions of arbitrary packages inside JavaScript bundles.
It fills the gap that production dependencies cannot be reliably determined through proxy metrics~\cite{Latendresse2022}.
At its core, Aletheia compares similarities through the rolling AST hash algorithm of Dolos~\cite{Maertens2022} and achieves a patch-level precision of 87\% for lab bundles and 82\% for real-world bundles (RQ2).
It clearly outperforms the previous BundlerStudy approach and is robust in the presence of first-party code and code reuse.
In addition, its file selection sub-algorithm successfully extracts relevant source code from npm release artifacts in 97\% of cases.
This facilitates our longitudinal study and increases accuracy and scale of future studies of the npm ecosystem.

Our Internet measurements attest CDN \glspl{url} a medium and bundles a high prevalence across the top 100,000 domains (RQ1).
Automatic updates provided through version aliasing are effective and responsible for most timely updates of \gls{cdn} resources.
Nonetheless, across all domains bundles show the timeliest update patterns with 17\% to 23\% of packages updated within 16 weeks (RQ3).
Bundles with associated pnpm source maps unveil to be a special subset of manually managed deployments with next to no updates in the first 4 weeks but a significant spike of updates afterwards.
Overall, these results highlight an extended period during which the vulnerabilities of the included packages may be exploited.
We observe 35\% of CDN-included and 2.7\% -- 3.7\% bundled packages per domain to contain a known vulnerability.
Furthermore, there are several indicators that few popular vendors have significant impact on the results.
In the future, these require special treatment to confirm that the more complex deployment process of bundles indeed promotes faster update patterns.

\bibliographystyle{ACM-Reference-Format}
\bibliography{bibliography.bib}

\end{document}